\documentclass[%
prb,twocolumn,amsmath,amssymb,superscriptaddress, bibnotes
]{revtex4-2}

\usepackage{graphicx}
\usepackage{bm}
\usepackage{color}
\usepackage{ulem}

\begin{document}

\title{Peak in the superconducting transition temperature\\ of the nonmagnetic topological line-nodal material CaSb$_2$ under pressure}

\author{Shunsaku~Kitagawa}

\email{kitagawa.shunsaku.8u@kyoto-u.ac.jp}
\affiliation{Department of Physics, Kyoto University, Kyoto 606-8502, Japan}

\author{Kenji~Ishida}
\affiliation{Department of Physics, Kyoto University, Kyoto 606-8502, Japan}

\author{Atsutoshi~Ikeda}
\affiliation{Department of Physics, Kyoto University, Kyoto 606-8502, Japan}
\affiliation{Department of Physics, University of Maryland, College Park, Maryland 20742-4111, USA}

\author{Mayo~Kawaguchi}
\author{Shingo~Yonezawa}
\author{Yoshiteru~Maeno}
\affiliation{Department of Physics, Kyoto University, Kyoto 606-8502, Japan}

\date{\today}

\begin{abstract}
Investigating the pressure dependence of the superconducting (SC) transition temperature $T_{\rm c}$ is crucial for understanding the SC mechanism.
Herein, we report on the pressure dependence of $T_{\rm c}$ in the nonmagnetic topological line-nodal material CaSb$_2$, based on measurements of electric resistance and alternating current magnetic susceptibility.
$T_{\rm c}$ initially increases with increasing pressure and peaks at $\sim$ 3.1~GPa.
With a further increase in pressure, $T_{\rm c}$ decreases and finally becomes undetectable at 5.9~GPa.
Because no signs of phase transition or Lifshitz transition are observed in the normal state, the peculiar peak structure of $T_{\rm c}$ suggests that CaSb$_2$ has an unconventional SC character.
\end{abstract}

\maketitle

The superconducting (SC) transition temperature $T_{\rm c}$ of a material is closely related to its electronic state.
In a conventional Bardeen-Cooper-Schrieffer (BCS) superconductor, $T_{\rm c}$ at the weak coupling limit is expressed as~\cite{J.Bardeen_PR_1957,pressureSC}
\begin{align}
T_{\rm c} = \frac{1.13\hslash \omega_{\rm D}}{k_{\rm B}}\exp\left(-\frac{1}{N(0)V}\right),
\label{eq.1}
\end{align}
where $\omega_{\rm D}$ is the Debye frequency, $\hbar$ is Dirac's constant, $k_{\rm B}$ is Boltzmann's constant, $N(0)$ is the density of states (DOS) at the Fermi energy $E_{\rm F}$, and $V$ is the effective interaction between the electrons mediated by the electron-phonon coupling.
Therefore, the $T_{\rm c}$ of a BCS superconductor is usually suppressed by the decrease in $N(0)$ owing to the bandwidth expansion when pressure is applied~\cite{T.F.Smith_PR_1967}.
In contrast, $T_{\rm c}$ exhibits a dome-like shape against tuning parameters, such as pressure and chemical substitution, when superconductivity is mediated by quantum critical fluctuations sensitive to these tuning parameters~\cite{K.Ishida_JPSJ_2009,C.Pfleiderer_RMP_2009}.
In addition, structural phase transitions or Lifshitz transitions change the electronic state dramatically, resulting in discontinuous variations in $T_{\rm c}$~\cite{T.C.Kobayashi_JPSJ_2011,S.Kitagawa_JPSJ_2013,A.Steppke_science_2017}.
Thus, investigating the pressure dependence of $T_{\rm c}$ is important for understanding the SC mechanism.

In this paper, we investigate the pressure dependence of $T_{\rm c}$ in CaSb$_2$ based on measurements of electric resistance and alternating current (AC) magnetic susceptibility.
We found that $T_{\rm c}$ shows a peak structure with a maximum value of 3.4~K at 3.1~GPa; this suggests the superconductivity of an unconventional nature in CaSb$_2$.

CaSb$_2$ crystallizes in a monoclinic structure with a non-symmorphic space group ($P2_1/m$, No. 11, $C_{2h}^2$). 
Based on band structure calculations \cite{K.Funada_JPSJ_2019}, CaSb$_2$ has Dirac line nodes in its bulk bands, which are protected by a combination of screw and mirror symmetries even with the spin-orbit coupling (SOC). 
In line-nodal materials, the intersections of two doubly degenerated bands, namely the Dirac point nodes, are connected to each other along a special line in the $k$-space.
These materials are predicted to exhibit several interesting phenomena, such as quasitopological electromagnetic response, owing to their special band structure~\cite{S.Ramamurthy_PRB_2017}.
Note that CaSb$_2$ has topologically nontrivial bands as well as trivial bands around $E_{\rm F}$.
Recently, some of the authors of the present study discovered superconductivity in this compound with $T_{\rm c} = 1.7$~K~\cite{A.Ikeda_PRM_2020}. 
CaSb$_2$ is expected to exhibit topologically nontrivial superconductivity owing to its topological bands.
However, at ambient pressure, $^{121/123}$Sb-nuclear quadrupole measurements reveal that the conventional metallic behavior originating from the Fermi-surface parts away from the nodes is dominant in the normal state and that the SC pairing symmetry is a conventional $s$-wave~\cite{H.Takahashi_JPSJ_2021}.
The variation in the SC properties with pressure will provide further insights regarding CaSb$_2$.

CaSb$_2$ were grown using the melt-growth method for resistance measurements (sample A) and the Sb self-flux method for AC susceptibility measurements (samples B--H).
The details of the melt-growth method were reported previously~\cite{A.Ikeda_PRM_2020}.
In the self-flux method, Ca and Sb in a molar ratio of 1:3.1 were placed in a tungsten crucible, which was sealed inside a quartz tube with Ar gas.
The tube was heated to 1000$^{\rm o}$C over 3 h, left undisturbed for 5 h, and then slowly cooled from 740$^{\rm o}$C to 600$^{\rm o}$C.
To identify the products, powder X-ray diffraction (XRD) was performed using a commercial diffractometer (Bruker AXS, D8 Advance) using Cu-K$\alpha$ radiation. 
In the melt-growth method, almost single-phase polycrystalline CaSb$_2$ was synthesized, whereas with the self-flux method, a mixture of small single crystals of CaSb$_2$ and Sb was obtained.
The superconductivity of Sb can be ignored because it appears only above 8.5~GPa ($T_{\rm c} \sim 3.5$~K)~\cite{J.Wittig_JPCS_1969}.
The energy dispersive x-ray analysis (EDX) spectra of CaSb$_2$ in all samples were measured using a commercial spectrometer (EDAX VE7800), and the composition ratio of Ca and Sb was determined to be approximately 1:2.
Pressure was applied using two types of pressure cells; an indenter-type pressure cell~\cite{T.C.Kobayashi_RSI_2007} was used for $P \leq$ 4~GPa and an opposed-anvil high-pressure cell, designed by Kitagawa $et$ $al.$~\cite{K.Kitagawa_JPSJ_2010}, was used for higher-pressure measurements.
Daphne 7575, which is liquid at an ambient condition and solidifies above 4 GPa,  was used as a pressure medium.
Based on the pressure dependence of $T_{\rm c}$ in a Pb manometer~\cite{A.Eiling_JPFMP_1981,B.Bireckoven_JPESI_1988}, we estimated the pressure value as follows:
\begin{align}
 P &= \frac{\Delta T_{\rm c}}{0.364} \hspace{15pt} \text{($P \leq$ 4~GPa)},\notag\\
 P &= \frac{\Delta T_{\rm c}}{0.364} + (\Delta T_{\rm c} - 1.456)^{1.6} \hspace{15pt} \text{($P >$ 4~GPa),}\notag
\end{align}
where $P$ is in GPa, and $\Delta T_{\rm c} = T_{\rm c}^{\rm Pb}(0) - T_{\rm c}^{\rm Pb}(P)$.
Electrical resistance was measured using a conventional four-terminal method with an AC resistance bridge (372, Lake Shore Cryotronics Inc.).
The AC magnetic susceptibility was measured using a self-inductance method.

The pressure dependence of $N(0)$ was calculated using the full-potential linearized augmented plane wave plus local orbitals method implemented in the {\sc WIEN2K} package~\cite{P.Blaha_wien2k_2018,P.Blaha_JCP_2020}.
We adopted the Perdew-Burke-Ernzerhof generalized gradient approximation~\cite{J.P.Perdew_PRL_1996} as the exchange-correlation functional.
We also considered the effect of SOC. 
The electronic states with varying lattice parameters were calculated.
We fitted the volume dependence of condensation energy with the Murnaghan equation of states and converted the volume change to pressure~\cite{F.D.Murnaghan_PNAS_1944}. 
The origin of the calculated pressure was set at the condensation-energy minimum at which the lattice parameters are slightly larger than the experimental values~\cite{K.Deller_casb2_1976}.
We then assumed two types of the pressure effect. 
One is isotropic compression, whereby, the unit cell shrinks without any variation in its aspect ratio. 
The other is $c$-axis compression.
For simplicity, we assume that only the $c$-axis lattice parameter shrinks and that the $a$- and $b$- axis lattice parameters do not change under pressure. 
Since CaSb$_2$ has a layered crystal structure, this assumption is more realistic even under hydrostatic pressure~\cite{Manikandan2020,A.A.Tsirlin_arXiv_2021}.

\begin{figure}[!tb]
\includegraphics[width=8.2cm,clip]{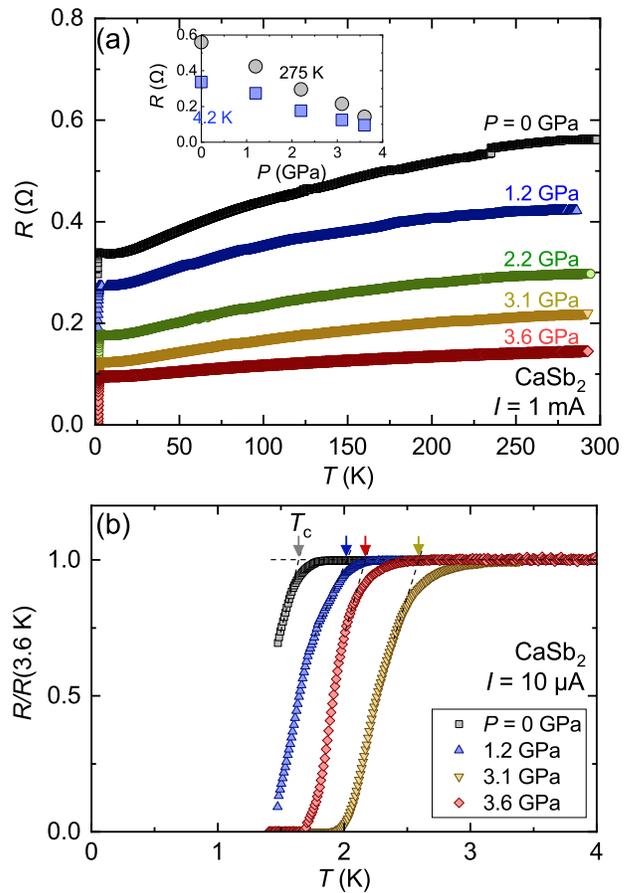}
\caption{(a) Temperature dependence of resistance $R$ at various pressures measured at 1~mA.
(Inset) Pressure dependence of $R$ at 4.2~K and 275~K.
(b) Temperature dependence of $R$ normalized by the value at 3.6~K measured at 10~$\mu$A.
$T_{\rm c}$, marked by arrows, was determined by the intersection of the slope of the resistance before and after the SC transition extrapolated by a straight line.
}
\label{Fig.1}
\end{figure}

Figure~\ref{Fig.1}(a) shows the temperature dependence of resistance $R$ at various pressures.
Because a sintered polycrystalline sample was measured and because its resistance value includes a large contribution from the grain boundary, it was difficult to estimate the exact value of resistivity.
At all pressures, $R$ decreased upon cooling, which is a typical behavior of a normal metal.
The value of $R$ linearly decreased with increasing pressure, as shown in the inset of Fig.~\ref{Fig.1}(a).
No other anomalies except for the SC transition were found in the pressure range up to 3.6~GPa.

We observed the SC transition at 1.7~K under ambient pressure.
Because of the limitations of the experimental conditions, the measurements were performed only down to 1.4~K.
Therefore, we defined the onset of the SC transition as $T_{\rm c}$, as shown in Fig.~\ref{Fig.1}(b).
$T_{\rm c}$ initially increased with increasing pressure and became 2.6~K at $\sim$ 3.1~GPa; this trend is opposite to that for conventional superconductors.
In contrast, $T_{\rm c}$ decreased with a further increase in pressure.

\begin{figure}[!tb]
\includegraphics[width=8.6cm,clip]{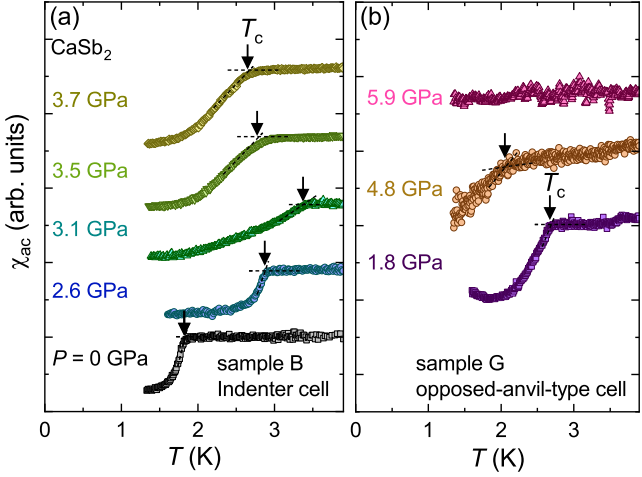}
\caption{Variation in the temperature dependence of the alternating current (AC) magnetic susceptibility with pressure for (a) sample B (measured using the indenter cell) and (b) sample G (measured using the opposed-anvil-type cell).
$T_{\rm c}$ was determined from the onset temperature and is indicated by the arrows.
}
\label{Fig.2}
\end{figure}

To check the nonmonotonic pressure dependence of $T_{\rm c}$, we also measured the AC magnetic susceptibility $\chi_{\rm ac}$, as shown in Fig.~\ref{Fig.2}.
$\chi_{\rm ac}$ shows a clear diamagnetic signal below 1.8~K at ambient pressure.
As in the resistance measurements, $T_{\rm c}$ was defined as the onset temperature of the SC transition.
The SC transition becomes broader under pressure due to the pressure distribution and the inhomogeneity of $T_{\rm c}$.
$T_{\rm c}$ exhibited a peak at $\sim$ 3.1~GPa, and no diamagnetic signal was observed down to 1.4~K at 5.9~GPa, which is consistent with the resistance measurements.

\begin{figure}[!tb]
\includegraphics[width=8.2cm,clip]{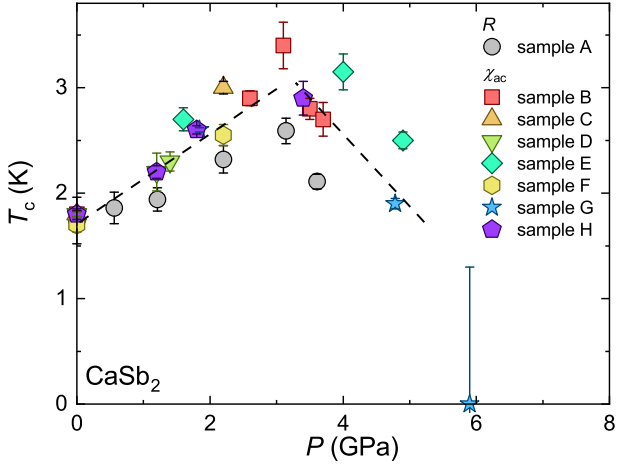}
\caption{Pressure--temperature phase diagram of CaSb$_2$.
The different symbols represent different samples.
$T_{\rm c}$ shows a peak at 3.1~GPa and was not observed at 5.9~GPa.
The dashed lines are intended for visual guidance.
The error bars for $R$ and $\chi_{\rm ac}$ were determined by the difference between the onset temperature and the temperature at which $R = 0.8R$(3.6~K) and the difference between the onset temperature and the temperature at which the resonance frequency shifts 50~kHz from the normal-state value, respectively.
The error bar at 5.9~GPa indicates out of measurement range.
}
\label{Fig.3}
\end{figure}

The pressure dependence of $T_{\rm c}$ is summarized in Fig.~\ref{Fig.3}.
The distribution of $T_{\rm c}$ could originate from that of sample quality.
In particular, the difference in the sample synthesis methods might be reflected in the difference in $T_{\rm c}$ between $R$ and $\chi_{\rm ac}$ measurements.
Nevertheless, all samples exhibited similar peak structures.
$T_{\rm c}$ initially increased with $dT_{\rm c}/dP~\sim~0.4$~K/GPa.
The maximum $T_{\rm c}$ was $\sim$ 3.4~K at 3.1~GPa in sample B, which is roughly double the $T_{\rm c}$ at ambient pressure.
Above 3.1~GPa, $T_{\rm c}$ decreased with $dT_{\rm c}/dP~\sim~-0.5$~K/GPa. 

\begin{figure}[!tb]
\includegraphics[width=8.2cm,clip]{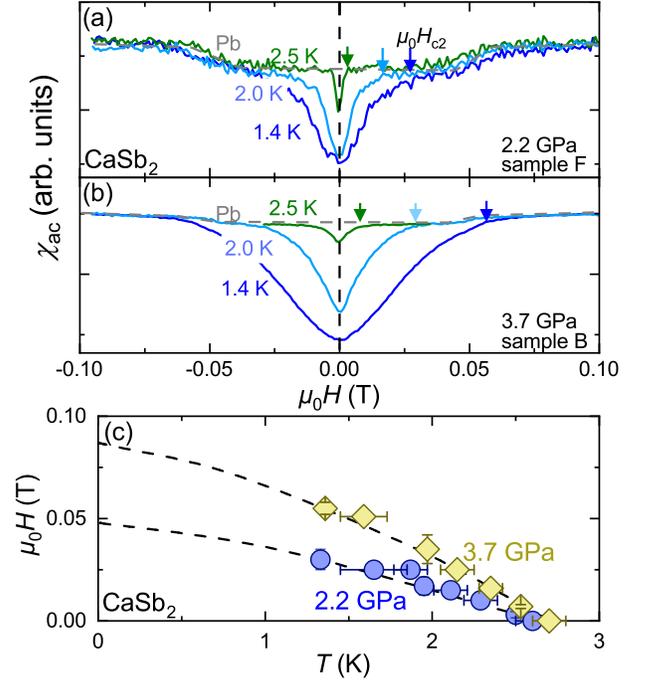}
\caption{
Magnetic field dependence of alternating current (AC) magnetic susceptibility at several temperatures measured at (a) 2.2~GPa and (b) 3.7~GPa.
The dashed curves represent the background signal from a Pb manometer.
$\mu_0 H_{\rm c2}$ was determined from the onset field of the superconducting diamagnetic signal from the sample and is indicated by the arrows.
(c)Temperature dependence of $H_{\rm c2}$ at 2.2 and 3.7~GPa determined based on AC magnetic susceptibility.
The dashed curves are intended for visual guidance.
The error bars were determined by the difference between the onset temperature/field and the temperature/field at which the resonance frequency shifts 50~kHz from the normal-state value, respectively.
}
\label{Fig.4}
\end{figure}

To investigate the change in the SC properties before and after the peak, the temperature dependence of the upper critical field $H_{\rm c2}$ at 2.2 and 3.7~GPa was also examined.
At these two pressures, $T_{\rm c}$ at zero field is almost the same ($\sim$ 2.7~K).
$H_{\rm c2}$ was determined from the magnetic field dependence of $\chi_{\rm ac}$ at a fixed temperature, as shown in Figs.~\ref{Fig.4}(a) and (b), and from the temperature dependence of $\chi_{\rm ac}$ at a fixed field, as shown in Fig.~S2.
Because we placed Pb inside the coil with the sample, the diamagnetic signal of Pb was also observed in both samples, as shown in Figs.~\ref{Fig.4}(a) and (b).
At both pressures, superconductivity was quickly suppressed by the magnetic field, and $H_{\rm c2}$ appeared to be governed by the orbital pair-breaking effect, as shown in Fig.~\ref{Fig.4}(c).
The estimated $\mu_0 H_{\rm c2}$(0)s at 2.2 and 3.7~GPa were 0.05~T and 0.08~T, respectively.
The change in $\mu_0 H_{\rm c2}$(0) is related to the change in the effective mass of the electron $m^{\ast}$.
The orbital critical field $\mu_0 H^{\rm orb}_{\rm c2}$ can be described as $\mu_0 H^{\rm orb}_{\rm c2} = 0.693(- d\mu_0 H_{\rm c2}/dT)_{T=T_{\rm c}} T_{\rm c} = \Phi_0/(2\pi \xi^2) = \pi \Delta^2_0 m^{\ast 2}/(2\hbar^4k^2_{\rm F})$~\cite{R.R.Hake_APL_1967}.
Here, $\Phi_0$ is the fluxoid quantum, $\xi = \hbar^2k_{\rm F}/\pi\Delta_0 m^{\ast}$ is the SC coherence length, $k_{\rm F}$ is the Fermi wave vector, and $\Delta_0$ is the SC gap.
$\mu_0 H_{\rm c2}$(0) at 3.7~GPa is 1.6 times larger than that at 2.2~GPa, suggesting a 1.3-fold increase in $m^{\ast}$.
Note that the $\mu_0 H_{\rm c2}$ of CaSb$_2$ is fairly small; it is even smaller than that of the type-I superconductor Pb.
However, because the lower critical field $\mu_0 H_{\rm c1}$ of CaSb$_2$ is $\sim$ 5~mT at ambient pressure~\cite{A.Ikeda_PRM_2020}, CaSb$_2$ is a type-II superconductor that is very close to a type-I superconductor, even under pressure.

\begin{figure}[!tb]
\includegraphics[width=8.2cm,clip]{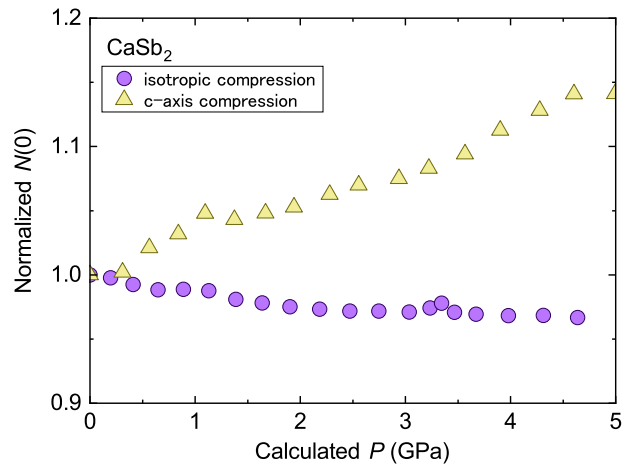}
\caption{Pressure dependence of the calculated $N(0)$ for isotropic and $c$-axis compression. 
$N(0)$ was normalized by the value at ambient pressure.
}
\label{Fig.5}
\end{figure}

Herein, we discuss the origin of the enhancement of $T_{\rm c}$ by applying pressure.
As mentioned above, $T_{\rm c}$ of BCS superconductors usually decreases with increasing pressure because of the decrease in $N(0)$~\cite{pressureSC}.
Indeed, as shown in Fig.~\ref{Fig.5}, the calculated $N(0)$ of CaSb$_2$ decreases slightly with isotropic compression and, at 3~GPa, it is approximately 97\% of the value at ambient pressure, while a small peak exists at 3.4~GPa.
However, there are some exceptions.
One is vanadium.
The $T_{\rm c}$ of vanadium monotonically increases with increasing pressure up to 120~GPa as a result of the suppression of spin fluctuations through the broadening of the $d$-band width~\cite{M.Ishizuka_PRB_2000}.
Based on our $^{121/123}$Sb-nuclear quadrupole measurements at ambient pressure~\cite{H.Takahashi_JPSJ_2021}, the SC pairing symmetry of CaSb$_2$ is a conventional $s$-wave, and conventional metallic behavior without low-energy spin fluctuations was observed in the normal state.
In addition, $N(0)V$ at ambient pressure was estimated to be $\sim 0.2$ from specific heat measurements~\cite{A.Ikeda_PRM_2020}, indicating a weak electron--phonon coupling.
In such a case, the pressure dependence of $V$ and $\omega_{\rm D}$ is weaker than that of $N(0)$ and is negligible~\cite{T.F.Smith_PR_1967}.
Therefore, it would be naively expected that the $T_{\rm c}$ of CaSb$_2$ decreases with the application of pressure; however, experimental results indicate the opposite.
One possible scenario to explain the experimental results is the increase in $N(0)$ with applying pressure due to anisotropic compressibility.
Actually, $N(0)$ at 3~GPa increases by approximately 10\% with $c$-axis compression, as shown in Fig.~\ref{Fig.5}, resulting in a $T_{\rm c}$ of 2.7~K, according to Eq.\eqref{eq.1}.
Thus, the experimentally determined $T_{\rm c}$ of 3.4~K cannot be explained only by the increase in $N(0)$ with $c$-axis compression. 
Furthermore, in reality, $N(0)$ may not increase so much because not only the $c$-axis but also the other crystal axes should shrink by hydrostatic pressure.
Therefore, an additional mechanism is necessary to completely understand the experimentally observed pressure dependence of $T_{\rm c}$.

The origin of the peculiar peak structure of $T_{\rm c}$ against pressure is also unknown.
In general, there are three possible reasons for a nonmonotonic change in $T_{\rm c}$.
The first is that the superconductivity is mediated by quantum critical fluctuations and appears near the quantum critical point~\cite{K.Ishida_JPSJ_2009,C.Pfleiderer_RMP_2009,Y.Kavashnin_PRL_2020,A.Majumdar_PRM_2020}.
In such a case, $T_{\rm c}$ exhibits a dome-like shape with respect to tuning parameters.
The second is that the electronic state is changed due to a phase transition.
A typical example is a drastic change in the electronic state due to a structural phase transition induced by pressure, resulting in a discontinuous change in $T_{\rm c}$~\cite{T.C.Kobayashi_JPSJ_2011,S.Kitagawa_JPSJ_2013}.
The final reason refers to the case of a change in the Fermi surface character passing through the van Hove singularity, namely the Lifshitz transition.
When a van Hove singularity exists, the DOS has a peak, resulting in a corresponding peak in the pressure dependence of $T_{\rm c}$~\cite{PhysRevB.1.214,A.Steppke_science_2017}.
There have been no reports of phase transitions in CaSb$_2$ other than the SC transition, and the measurements in this study did not indicate such a transition.
In addition, the absolute values of the slope of $T_{\rm c}$ with respect to pressure as well as $H_{\rm c2}$ are almost the same before and after the peak of $T_{\rm c}$, suggesting the absence of a substantial change in the electronic state.
Therefore, the possibility of the first or second reason is marginal.
Moreover, the last reason is inconsistent with the result of the band calculations.
As shown in Fig.~\ref{Fig.5}, $N(0)$ almost monotonically decreases(increases) with isotropic compression($c$-axis compression) instead of peaking at $\sim$ 3~GPa.
The origin of the peak structure in $T_{\rm c}$ is therefore unknown.

The remaining possibility is that the complex pressure dependence of the lattice parameters results in the peak structure of $N(0)$.
More detailed investigations, including XRD measurements under pressure and microscopic measurements, will clarify this behavior.
Since our measurements may have overlooked phase transitions that exist, it is necessary to measure high-quality single-crystal samples.
Furthermore, it is important to investigate the SC symmetry in the high-pressure region because the SC symmetry above 3.1~GPa could be different from the $s$-wave realized at ambient pressure. 
It was theoretically proposed that an SC symmetry other than an $s$-wave ($A_g$ in the irreducible representation of $C_{2h}$) in CaSb$_2$ should be topologically nontrivial~\cite{S.Ono_arXiv_2020}.

In conclusion, we observed a peculiar pressure dependence of the $T_{\rm c}$ of CaSb$_2$ from the measurements of electrical resistance and AC magnetic susceptibility.
The maximum $T_{\rm c}$ was 3.4~K at 3.1~GPa, which is approximately double that at ambient pressure.
Because there are no signs of phase transition or Lifshitz transition up to 4~GPa in the normal state, the peculiar peak structure in $T_{\rm c}$ suggests that the SC character of CaSb$_2$ is unconventional and changes at around 3~GPa.
To confirm the origin of this peak structure and the SC mechanism under high pressure, further investigations are desired.
Our findings provide a new perspective for investigating the SC properties of topologically nontrivial materials. 

{\it Note added. After submitting this Letter, we noticed that the superconducting properties of single crystalline CaSb$_2$ at ambient pressure have been reported by another group~\cite{M.Oudah_arXiv_2021}.}

\section*{Acknowledgments}
The authors thank H. Takahashi for valuable discussions. 
This work was partially supported by Kyoto University LTM center, Grant-in-Aid for Scientific Research on Innovative Areas from the Ministry of Education, Culture, Sports, Science, and Technology (MEXT) of Japan, and JSPS Core-to-core program (No. JPJSCCA20170002). 
We are also supported by JSPS KAKENHI Nos. JP15H05852, JP15K21717, JP17H06136, JP20H00130, JP15K21732, JP15H05745, JP20KK0061, JP19H04696, JP19K14657, and JP20H05158.

\end{document}